\documentclass[fleqn,usenatbib]{mnras}

\usepackage{newtxtext,newtxmath}
\usepackage[T1]{fontenc}

\usepackage{xcolor}
\usepackage{graphicx}	
\usepackage{amsmath}	
\usepackage{newtxtext,newtxmath}
\usepackage{nicefrac}
\usepackage{tikz}       
\usepackage[mathscr]{eucal}  
\usepackage[smallscripts]{moresize} 
\usepackage[outline]{contour} 

\newcommand{\fett}[1]{\boldsymbol{#1}}
\newcommand{\dd}{{\rm{d}}}
\newcommand{\ii}{{\rm{i}}}
\newcommand{\be}{\begin{equation}}
\newcommand{\ee}{\end{equation}}
\newcommand{\nabx}{\boldsymbol{\nabla}_{\boldsymbol{x}}}
\newcommand{\nabq}{\boldsymbol{\nabla}_{\boldsymbol{q}}}
\newcommand{\nab}{\fett{\nabla}}
\newcommand{\mD}{\text{\small $D$}}
\newcommand{\cE}{c_{\hskip-0.02cm\text{\tiny $E$}}}
\newcommand{\cF}{c_{\hskip-0.02cm\text{\tiny $F$}}}
\newcommand{\pD}{\partial_{\text{\fontsize{7}{7}\selectfont\it D}}}
\newcommand{\mB}{\text{\ssmall B}}
\newcommand{\mDirac}{\text{\ssmall D}}
\newcommand{\mthree}{\text{\ssmall $(3)$}}
\newcommand{\pDsq}{\partial_{\fontsize{7}{7}\selectfont\it D}^2}
\newcommand{\mPsi}{{\mathit \Psi}}
\newcommand{\sdfrac}[2]{\mbox{\small$\displaystyle\frac{#1}{#2}$}}
\newcommand{\ws}{\hspace{0.02cm}}

\definecolor{lime}{HTML}{A6CE39}
\DeclareRobustCommand{\orcidicon}{
	\begin{tikzpicture}
	\draw[lime, fill=lime] (0,0) 
	circle [radius=0.14] 
	node[white] {{\fontfamily{qag}\selectfont \tiny ID}};
	\draw[white, fill=white] (-0.0625,0.095) 
	circle [radius=0.007];
	\end{tikzpicture}
	\hspace{-2mm}
}

\foreach \x in {A, ..., Z}{%
	\expandafter\xdef\csname orcid\x\endcsname{\noexpand\href{https://orcid.org/\csname orcidauthor\x\endcsname}{\noexpand\orcidicon}}
}

\contourlength{0.3pt} 
\contournumber{6} 

\title[$\mathit{\Lambda}$CDM growth functions] {Analytical growth functions for cosmic structures in a  \contour{black}{$\Lambda$}CDM Universe}

\author[Rampf et al.]{
 Cornelius Rampf,$^{{\,{\tiny\orcidA{}}\,\,\,\,1,2}}$\thanks{E-mail: cornelius.rampf@univie.ac.at}  
 Sonja Ornella Schobesberger$^{{\,{\tiny\orcidB{}}\,\,\,\,1,2}}$
and Oliver Hahn$^{{\,{\tiny\orcidC{}}\,\,\,\,1,2}}$
\\
${}^1$Department of Astrophysics, University of Vienna, Türkenschanzstraße 17, 1180 Vienna, Austria \\ 
${}^2$Department of Mathematics, University of Vienna, Oskar-Morgenstern-Platz 1, 1090 Vienna, Austria
}

\date{Accepted XXX. Received YYY; in original form ZZZ}

\begin{document}
\label{firstpage}
\pagerange{\pageref{firstpage}--\pageref{lastpage}}
\maketitle

\begin{abstract}
The cosmological fluid equations describe the early gravitational dynamics of cold dark matter (CDM), exposed to a uniform component of dark energy, the cosmological constant $\Lambda$. Perturbative predictions for the fluid equations typically assume that the impact of $\Lambda$ on CDM can be encapsulated by a refined growing mode $D$ of linear density fluctuations. Here we solve, to arbitrary high perturbative orders, the nonlinear fluid equations with an {\it Ansatz} for the fluid variables in increasing powers of~$D$. We show that $\Lambda$ begins to populate the solutions  starting at the fifth order in this strict $D$-expansion. By applying suitable resummation techniques, we recast these solutions to a standard perturbative series where not $D$, but essentially the initial gravitational potential serves as the bookkeeping parameter within the expansion. Then, by using the refined growth functions at second and third order in standard perturbation theory, we determine the matter power spectrum to one-loop accuracy as well as the leading-order contribution to the matter bispectrum. We find that employing our refined growth functions impacts the total power- and bispectra at a precision that is below one percent at late times. However, for the power spectrum, we find a characteristic scale-dependent suppression that is fairly similar to what is observed in massive neutrino cosmologies. Therefore, we recommend employing our refined growth functions in order to reduce theoretical uncertainties for analysing data in related~pipelines. 
\end{abstract}


\section{Introduction}

Analytical predictions for the cosmic large-scale structure are essential in various aspects related to cosmological observations, such as for accurately extracting the baryonic oscillation feature \citep{2011MNRAS.416.3017B,2011MNRAS.418.1707B,2018MNRAS.474.2109S}, or more generally for inferring the cosmological model from observations by employing various summary statistics \citep[e.g.][]{1998ApJ...495...44C,2012MNRAS.420...61K,2012JCAP...11..029G,2014JCAP...01..042D,2018JCAP...04..030S,2020JCAP...11..035C}.
Reliable theoretical predictions are also important for accurate forward modelling at the field level \citep{2015arXiv151204985L}, in particular also within the context of effective fluid descriptions \citep{2019JCAP...01..042S,2021JCAP...08..029B,2021JCAP...04..033S}.

The main theoretical ingredient of such avenues is cosmological perturbation theory (PT), which delivers analytical solutions based on the cosmological fluid equations for cold dark matter (CDM);  see e.g. \cite{1980lssu.book.....P,Bernardeau2002} for related textbooks or reviews, and e.g.\ \cite{2006PhRvD..73f3519C,2008JCAP...10..036P,2013MNRAS.429.1674C,2014JCAP...01..010B,2015PhRvD..91b3508V} for rather recent PT findings. 
Historically, the first PT solutions assumed a simplified cosmological set-up that ignored the gravitational coupling between CDM and the cosmological constant $\Lambda$ (and assumed zero spatial curvature), which is the so-called Einstein--de Sitter (EdS) cosmological model.

Perturbative solutions that incorporated $\Lambda$ were first derived at the linear level, and expressed in terms of exact (hypergeometric) functions by \cite{1977MNRAS.179..351H,1992ARA&A..30..499C,1992A&A...263...23B,2003A&A...399...19C}, while an analytical second-order solution is given by \cite{1995PThPh..94.1151M}. Beyond second order,  we are not aware of exact analytical solutions; but see \cite{1995A&A...296..575B} for asymptotic considerations leading to third-order solutions.

In the sense of standard PT within the Newtonian limit, the initial gravitational potential $\varphi^{\rm ini}$ can be thought of as the book keeping parameter within the expansion. Thus, the $n$th-order perturbative solution is proportional to $\sim (\varphi^{\rm ini})^{n}$, modulo various spatial derivatives acting on these fields.
Furthermore, in the 90s, it was realised that the $n$th-order perturbative solution is proportional to $D^n$ to a ``very good'' approximation (considering  the target accuracy at these times), where $D$ (or $D_+$) is the linear growing mode of linear density fluctuations. It is the purpose of the present paper to systematically analyse the precise relation between a standard expansion and a strict $D$ expansion.

Perturbative calculations can be performed in either a fixed Eulerian coordinate system, or in a co-moving system that employs Lagrangian (initial) coordinates. For most parts of the present paper, we employ Lagrangian coordinates which has certain advantages as opposed to the Eulerian way. Just to name a few, perturbative solutions in Lagrangian coordinates, usually dubbed Lagrangian perturbation theory \citep[LPT; e.g.][]{1989A&A...223....9B,1991ApJ...382..377M,1992ApJ...394L...5B,1992MNRAS.254..729B,1995A&A...296..575B,1997GReGr..29..733E,2012JCAP...06..021R,2014JFM...749..404Z},  can be straighforwardly applied to generate initial conditions for cosmological $N$-body simulations \citep{1983MNRAS.204..891K,1985ApJS...57..241E,1998MNRAS.299.1097S,2006MNRAS.373..369C,2021MNRAS.500..663M,2021MNRAS.503..426H}.
Secondly, in Lagrangian coordinates it is fairly straightforward to detect the breakdown of the standard perturbative framework, which is linked to the crossing of particle trajectories \citep{2017MNRAS.471..671R,2018PhRvL.121x1302S,2021MNRAS.501L..71R,2022A&A...664A...3S}. Thirdly, 
Lagrangian coordinates are optimal to describe advection of perfectly cold fluids and, as a consequence, LPT solutions generically converge much faster than the PT solutions in Eulerian coordinates at fixed truncation order (but see \citealt{2019PhRvD..99h3524U} for an exception within a semi-classical context).

This paper is organised as follows. In section~\ref{sec:setup}, we provide the basic equations governing the CDM evolution, formulated in Lagrangian space where the whole dynamical information of the system is encapsulated in the Lagrangian displacement field.
In section~\ref{sec:Displacement} we then derive novel all-order recursive relations for the displacement in terms of the linear growing mode $D$ in a $\Lambda$CDM Universe.
This analytical form of recursive relations allows us to recast the results within the $D$ expansion into a pertubative series in standard form; see sections~\ref{sec:prelimGrowth}--\ref{sec:resum}. After discussing normalisation and truncation effects in section~\ref{sec:norm-trunc}, we analyse our expressions for the structure growth and velocity, and compare them against related findings in the literature in section~\ref{sec:discussgrowth} (see Appendix~\ref{app:disp-velocity} for a concise list of results). 
Finally, we determine the one-loop matter power spectrum and tree-level matter bispectrum in section~\ref{sec:polyspectra}, and conclude with a summary and discussions in section~\ref{sec:concl}.

\section{Setup} \label{sec:setup}

\subsection{Basic equations}

We denote by  $\fett{u} = \dot{\fett{x}}$ the peculiar velocity, where a
dot denotes the (convective) time derivative with respect to cosmic time $t$,
while $\fett{x} = \fett{r}/a$ are comoving coordinates and $a$ is the cosmic scale factor.
The equations of motion for collisionless matter elements in a $\Lambda$CDM Universe are \citep{1980lssu.book.....P}
\be  \label{eq:EOMbasic}
  \ddot{\fett{x}} + 2H \dot{\fett{x}} = - \tfrac{1}{a^2}  \nabx \phi\ , \qquad \quad   \nabx^2 \phi = 4\uppi G \bar \rho a^2 \delta(\fett{x})
 \,,
\ee
where  $H = \dot a/a$ is the Hubble parameter, $\bar \rho \propto a^{-3}$ the mean matter density, and $\delta = (\rho -\bar \rho)/\bar \rho$ the density contrast.

Taking mass conservation into account and formally linearising the equations of motion, one arrives at the
standard ordinary differential equation for the linear density contrast $\delta = \delta_1$,
\be \label{eq:ODElindelta}
 \ddot \delta_1 + 2H \dot \delta_1 = 4\uppi G \bar \rho\, \delta_1 \,.
\ee
This well-known equation has two solutions, one is decaying as $\sqrt{1+\Lambda a^3}a^{-3/2}$, and the other is the growing-mode solution,
which is usually denoted with $D$ or $D_+$, and 
can be expressed in terms of the hypergeometric function ${}_2F_1$  \citep{2003A&A...399...19C},
\be \label{eq:Dplus}
  D(a)\! = a \sqrt{1+ \Lambda a^3} \,{}_2F_1\!\left(\tfrac 3 2, \tfrac  5 6,\tfrac{11}{6}, - \Lambda a^3 \right) =  a - \frac{2\Lambda}{11}a^4 + O(a^7).
\ee 
Here, we have defined $\Lambda = \Omega_{\Lambda0}/\Omega_{\rm m0}$, where 
$\Omega_{\rm m 0}$ and $\Omega_{\Lambda0}$ are respectively the present-day values of the matter density and cosmological constant at $a=a_0$. 
Without loss of generality we set $a_0=1$, as well as $\Omega_{\rm m0} =0.302$ and $\Omega_{\Lambda0} =0.698$ \citep{2020A&A...641A...6P}.
We note that it is customary in the literature to also normalise $D$ such that $D( \text{\small $a_0=1$})=1$, which is {\it not} yet applied in~\eqref{eq:Dplus}. In what follows we keep the above relationship to avoid unnecessary cluttering, but  in section~\ref{sec:norm-trunc}
we provide simple instructions on how to adapt to the standard normalisation to our final results (see Appendix~\ref{app:disp-velocity} for normalised results).

In this paper we solve equations~\eqref{eq:EOMbasic} by means of a Taylor expansion in powers of the linear growing mode $D$. For this it is essential to convert all temporal dependences in~\eqref{eq:EOMbasic} from $t$-time to $D$-time, which implies for the involved temporal derivatives
\be \label{eq:deriv}
 \partial_t = \dot D \pD , \quad\, \partial_t^2 = \partial_t (\dot D \pD) = \left[ 4 \uppi G \bar \rho D - 2H \dot D \right] \pD   + \dot D^2 \pDsq ,
\ee
where we used that $\ddot D = 4\uppi G \bar \rho D - 2H \dot D$ which follows from Eq.\,\eqref{eq:ODElindelta}.
After replacing the temporal derivatives as instructed through~\eqref{eq:deriv}, equations~\eqref{eq:EOMbasic} can be equivalently written as
\be \label{eq:master}
  \left[ 1 - \lambda(\mD) \right] \pDsq \fett{x} + \frac{3}{2D} \pD \fett x  = - \frac{3}{2D} \nabx \varphi \,, \qquad \,\, \nabx^2 \varphi = \frac \delta D \,.  
\ee
Here, we employ the rescaled potential
$\varphi \equiv \phi/(4\uppi G \bar \rho a^2 D)$,
and we have defined 
\be
  1- \lambda(\mD) = \left[ 1+ \Lambda a^3 \right] \left( \frac{a \partial_a D}{D} \right)^2 \,,
\ee
which is precisely equal to the quantity $f^2/\Omega_{\rm m}$ frequently encountered in the literature, with $f = \dd \ln D / \dd \ln a$ and $\Omega_{\rm m} =8\uppi G \bar \rho/(3H^2)$.
In deriving equation~\eqref{eq:master} we have made use of the first Friedmann equation.
Note that the only explicit cosmological dependence in~\eqref{eq:master}
is through the occurring  ``$\lambda$'' term --  a property first shown by \cite{1998ApJ...496..586S}.
Furthermore, since the series representation of $D(a)$ can be inverted 
to $a(\mD) = D + (2\Lambda/11)D^4 + O(D^7)$,
it follows that~$\lambda$ can also be represented as a series in~$D$,
\begin{subequations} \label{eqs:lambda}
\be \label{eq:lambda}
  \lambda(\mD) = \sum_{n=1}^\infty \lambda^{(n)} D^n  \,,
\ee
where the first non-vanishing coefficients are
\be \label{eq:lambdas}
 \lambda^{(3)} = \frac \Lambda{11}\,, \qquad \lambda^{(6)} =  \frac{30 \Lambda^2}{2057} \,,  \qquad \lambda^{(9)} =  \frac{1560 \Lambda^3}{520\,421}\, .
\ee
\end{subequations}
In the limit of a spatially flat, matter dominated universe with vanishing cosmological constant, we have $\lambda \to 0$ and $D \to a$, and equations~\eqref{eq:master}  coincide with the standard equations for an EdS model \citep[see e.g.][]{Bernardeau2002}.

\subsection{Equations for the Lagrangian displacement field}

To proceed, a few standard operations are necessary to arrive at our evolution equations in ready-to-use form.
We begin by taking the Eulerian divergence of the first equation in~\eqref{eq:master}, which allows us to write its right-hand-side in terms of the Poisson equation. To close the system of equations, we also demand the vanishing of the Eulerian vorticity $\nabx \times \fett u = \fett 0$.
As an intermediate stage, we then have
\be \label{eq:mainSemiLag}
  \nabx \cdot \left\{ [1-\lambda] \pDsq + \frac{3}{2D} \pD \right\} \fett x = - \frac{3\delta}{2D^2}\,, \qquad  
 \nabx \times  \pD{\fett x} = \fett 0   \,.
\ee
Furthermore, we employ the Lagrangian displacement field~$\fett{\mPsi}$ which encapsulates
the trajectories of fluid elements from initial position~$\fett q$ (at $D=0$) to current/Eulerian position~$\fett x$ at time~$D$,
\be
  \fett{x}(\fett{q},\mD) = \fett{q}+ \! \fett{\mPsi}(\fett{q},\mD) \,.
\ee
As long as the dark-matter fluid is single stream, the fluid density is controlled by the Jacobian $J$ of the mapping $\fett q \mapsto \fett x$,
\be
 \delta(\fett x(\fett q, \mD), \mD) = \frac 1 J -1 \,, \qquad\, \, J = \det (\delta_{ij}+ \mPsi_{i,j})  \,.
\ee
Here, $\delta_{ij}$ is the Kronecker delta, Latin indices denote the three Cartesian components, and~``$,j$'' is a partial derivative  with respect to Lagrangian component~$q_j$.

Finally, we convert the  Eulerian space derivatives in equation~\eqref{eq:master} 
to Lagrangian derivatives for which we use the identity 
\mbox{$\nabla_{x_i}=  (\partial q_j/\partial x_i)  \nabla_{q_j} = (2J)^{-1} \varepsilon_{ikl}\varepsilon_{jmn} x_{k,m}x_{l,n} \nabla_{q_j}$}, 
where summation over repeated indices is assumed and $\varepsilon_{ikl}$ denotes the Levi--Civit\`a symbol. With these relations, it is easily shown that equations~\eqref{eq:mainSemiLag} can be equivalently written as
\be \label{eq:LagFull}
  \boxed{ \hat {\cal T}_{1} \, \Big\{ \nab \cdot  \! \fett \mPsi \Big\} = {\cal W}\, , \qquad \,\,
  \pD \Big\{ \nab \times \! \fett \mPsi \Big\} =  \nab   \mPsi_l \times \nab  \pD \mPsi_l  \, , }
\ee
where $\nab \equiv \nabq$ from here on, and we have defined the operator 
\be \label{eq:Top}
 \hat {\cal T}_n :=  \nicefrac 2 3  [1-\lambda] D^2 \pDsq + D\ws\pD -n \,,
\ee
as well as the non-linear source term
\begin{align}
  {\cal W} = [  \mPsi_{j,i} - \mPsi_{l,l} \delta_{ij} ]\, \hat {\cal T}_{\nicefrac 1 2\,} \mPsi_{i,j} - \tfrac 1 2 \varepsilon_{ikl} \varepsilon_{jmn} \mPsi_{k,m} \mPsi_{l,n} \hat  {\cal T}_{\nicefrac 1 3\,} \mPsi_{i,j} .
\end{align}
Equations~\eqref{eq:LagFull} are the main evolution equations that we solve in this paper.
We remark that variants of these equations are well-known in the literature \citep[see e.g.][]{1995A&A...296..575B,1997GReGr..29..733E,2015PhRvD..92b3534M}, but, except for~\cite{2021MNRAS.503..406R,2021RvMPP...5...10R}, they are  not  formulated exclusively in $D$-time which is however crucial for the present considerations.

\section{Displacement and its time-Taylor series} \label{sec:Displacement}

\subsection{Linear analysis in Lagrangian space} \label{sec:ZA}

Formally linearising around the  background-solution
$\!\fett \mPsi =0$ in the evolution equations, all terms on the right-hand-sides of~\eqref{eq:LagFull}  are negligible, thus leading to
\be \label{eq:linearODEs}
   \hat {\cal T}_{1} \, \Big\{ \nab \cdot \! \fett \mPsi \Big\} = 0 , \qquad \quad\,
  \pD \Big\{ \nab \times \!\fett \mPsi \Big\} =  0 ,
\ee
with the solutions 
\be
  \nab \cdot \!  \fett \mPsi = C_+ D + C_- \mbox{$\sqrt{1+ \Lambda a^3(\mD)}$} \, a^{-3/2}(\mD) , 
  \quad \,\, \nab \times \! \fett \mPsi  = \fett{w}_0  ,
\ee
respectively,
where $C_+, C_-$ and $\fett w_0$ are spatial constants, and we remind the reader 
that $a(\mD) = D + (2\Lambda/11)D^4 + O(D^7)$. The term $\propto C_-$ is decaying
as $D^{-3/2}$ for short times and is vastly suppressed at later times in comparison to the term $\propto C_+$, hence the customary wording ``decaying term'' in the literature. 
Furthermore, a nonzero~$\fett w_0$ would signal the presence of initial vorticity which is however usually not assumed \citep[but see e.g.][]{1992MNRAS.254..729B,2016PhRvD..94h3515R}.
It was shown by \cite{Brenier:2003xs}
that suitable boundary conditions can be employed such that only the purely growing-mode and vorticity-free solutions are selected \citep[see also][]{2014JFM...749..404Z,2015MNRAS.452.1421R}. We follow their approach which implies demanding $\delta^{\rm ini} = 0$ and $\pD \fett{x}|_{D=0} \equiv \fett v^{\rm ini} = - \fett \nabla \varphi^{\rm ini}$  at initial time $D=0$ (denoted with superscript ``ini'').
Within this set-up, one obtains from~\eqref{eq:linearODEs} the well-known Zel'dovich solution with corresponding trajectory
\be \label{eq:ZA}
  \fett x (\fett q, \mD) = \fett q + D \, \fett v^{\rm ini} \,, \qquad \qquad \fett v^{\rm ini}(\fett q) = - \fett \nabla \varphi^{\rm ini} \,.
\ee
See \cite{2021MNRAS.500..663M} for explicit instructions about how to obtain the initial gravitational potential $\varphi^{\rm ini}$, and e.g.\ \cite{1992MNRAS.254..729B,1995A&A...296..575B,2013MNRAS.431..799N} for incorporating decaying modes in the analysis.

\subsection{Time-Taylor series for the non-linear displacement} 

In the standard approach of LPT, the system of equations~\eqref{eq:LagFull} is solved in a perturbative expansion in increasing powers of small displacements (or, equivalently, in powers of $\varphi^{\rm ini}$). Here we pursue a somewhat different strategy, that is we seek a {\it Taylor series representation} of the non-linear displacement field, where the refined time variable, the linear growth $D$, acts as the expansion parameter, 
\be \label{eq:psiansatz}
  \fett \mPsi (\fett q, \mD) = \sum_{n=1}^\infty \fett{\psi}^{(n)}(\fett q) \, D^n \,,
\ee
where $\fett{\psi}^{(1)}(\fett q)  = - \nab \varphi^{\rm ini}$ is the spatial Taylor coefficient corresponding to the Zel'dovich solution~\eqref{eq:ZA}. A few comments are in order. Firstly, 
it is clear that this {\it Ansatz} only selects the purely growing-mode solutions in accordance with the chosen boundary conditions (see text before equation~\ref{eq:ZA}).
Secondly, for the limiting case of an EdS cosmological model where $\Lambda =0$ and $D=a$, the perturbative results in standard LPT  agree exactly with those obtained through the above {\it Ansatz} \citep[cf.][]{2015MNRAS.452.1421R}.
Thirdly and most importantly, in a $\Lambda$CDM Universe where $\Lambda$ appears explicitly in the evolution equations~\eqref{eq:LagFull}, fixed-order results obtained from the time-Taylor expansion~\eqref{eq:psiansatz} will generally differ from standard perturbative results.  
None the less, in section~\ref{sec:LSSGrowth} we show how our 
time-Tayor results can be converted to those of the standard approach.

For $n>1$, it is by now a standard procedure to determine the unknowns~$\fett{\psi}^{(n)}(\fett q)$ in~\eqref{eq:psiansatz}. In the following we provide the key steps and refer to
 e.g.~\cite{2014JFM...749..404Z,2015MNRAS.452.1421R,2015PhRvD..92b3534M}  for in-depth derivations applied to fairly analogous set-ups. Plugging the {\it Ansatz}~\eqref{eq:psiansatz} into the evolution equations~\eqref{eq:LagFull}, one identifies the coefficients of the various powers of~$D$.
For this it is important to observe that terms~$\propto \lambda$ in~\eqref{eq:LagFull} constitute in general higher-order contributions in comparison to 
terms that are not decorated with~$\lambda$'s.
 After appropriate symmetrisation of the various terms, we find recursive relations for the longitudinal and transverse parts of~$\fett \psi^{(n)}$, which are respectively denoted with~$\mu_1^{(n)} \equiv \psi_{l,l}^{(n)}$ and~$A_i^{(n)} \equiv \varepsilon_{ijk} \,\psi_{k,j}^{(n)}$. The resulting all-order recursion relations are
\begin{subequations}
\label{eq:recs}
\begin{align}
  & \mu_1^{(n)} \!=   - \varphi_{,ll}^{\rm ini} \delta_1^n    
  + \!\!  \sum_{i+j = n} \!\!\left[ \tfrac{( 3 - n)/ 2 - i^2 - j^2}{(n+ 3/2) (n-1)} \mu_2^{(i,j)} + \tfrac{i(i-1) }{(n+ 3/2) (n-1)} \mu_1^{(i)} \lambda^{(j)}  \right] \nonumber \\
 &\quad\,\,\ws + \sum_{i+j+k = n}  \left[ \tfrac{(3 -n) /2 - i^2 -j^2 -k^2}{(n+ 3/2) (n-1)} \mu_3^{(i,j,k)}   
      + \tfrac{i(i-1) + j(j-1)}{(n+3/2) (n-1)}  \mu_2^{(i,j)} \lambda^{(k)} \right]  \nonumber \\
&\quad\,\, \ws + \sum_{i+j+k+l =n} \tfrac{i(i-1) + j(j-1)  + k(k-1)}{(n+3/2) (n-1) } \mu_3^{(i,j,k)} \lambda^{(l)} , \label{eq:recLongitudinal} \\ 
  &A_i^{(n)} = 
   \sum_{0 < s <n}  \tfrac{n-2s}{2n} \, \varepsilon_{ijk} \,\psi_{l,j}^{(s)}  \psi_{l,k}^{(n-s)}  \ , \label{eq:recTrans}
\end{align}
\end{subequations}
where the $\lambda^{(i)}$ coefficients are given in~\eqref{eqs:lambda}.
In the above equations we have also employed the purely spatial functions 
\begin{subequations} \label{eqs:mu2andmu3}
\begin{align}
  &\mu_2^{(n_1,n_2)}  = \frac{1}{2} \left[  \psi_{i,i}^{(n_1)} \psi_{j,j}^{(n_2)} -  \psi_{i,j}^{(n_1)}  \psi_{j,i}^{(n_2)} \right]\,, \\ 
  &\mu_3^{(n_1,n_2,n_3)}  = \frac{1}{6} \varepsilon_{ikl}\varepsilon_{jmn}\psi_{k,m}^{(n_1)}\psi_{l,n}^{(n_2)}\psi_{i,j}^{(n_3)} \,,
\end{align}
\end{subequations}
and we note that all $\mu$ and $\lambda$ terms appearing in~\eqref{eq:recs} are zero if any of its perturbation indices are zero or negative.
Equations~\eqref{eq:recs} allow us to determine the longitudinal and transverse part of the displacement coefficient $\fett \psi^{(n)}$ at successively high orders, once the initial gravitational potential is fixed.
For this one considers the Helmholtz decomposition 
\be
  \fett{\psi}^{(n)} = \fett \nabla^{-2} \left( \nab \mu_1^{(n)} - \nab \times \fett{A}^{(n)} \right)\,,
\ee 
from which $\!\fett{\mPsi}(\fett q, \mD) = \sum_n \fett{\psi}^{(n)}(\fett q)\,D^n$ readily follows.

\section{Structure growth in perturbation theory} \label{sec:LSSGrowth}

\subsection{Preliminary considerations} \label{sec:prelimGrowth}

One important feature of the recursive relations~\eqref{eq:recs} is the explicit appearance of $\Lambda$ through the $\lambda^{(i)}$'s. Specifically, it is easily checked that $\Lambda$ begins to populate the infinite displacement series at order~$D^5$:
\be
  \nab \cdot \! \fett \mPsi  \supset - \nab^2 \varphi^{\rm ini} D - \sdfrac 3 7 \mu_2^{(1,1)}(\fett q)\, D^2
    - \sdfrac{3\Lambda}{1001} \mu_2^{(1,1)} (\fett q)\, D^5 .
\ee
Here it is essential to observe that both the contributions $\propto D^2$ and $\propto D^5$  share a common spatial term, $\mu_2^{(1,1)}$. 
Furthermore, going to consecutive higher powers in~$D$, one may expect  an infinite tower of $\Lambda$-contributions proportional to $\mu_2^{(1,1)}$ which, ultimately, would alter the temporal growth of that spatial term. A straightforward inspection shows that this is indeed the case, albeit with decreasing significance at higher orders, as we will see.

Of course, the above argument is not only valid for terms $\propto \mu_2^{(1,1)}$ but applies to the temporal growth of all involved spatial terms (except the linear displacement which is precisely proportional to $D$).
Consider the following standard perturbative expansion of the Lagrangian displacement where one introduces the expansion
 parameter $\epsilon >0$, which essentially keeps track of the involved powers  in the initial gravitational potential \citep[or initial matter fields; e.g.][]{1995A&A...296..575B,1995PThPh..94.1151M}, i.e.,
\begin{align} 
 &\!{\fett{\mPsi}} = \epsilon D  \fett{\mPsi}^{(1)}(\fett q)+ \epsilon^2 E(\mD) \fett{\mPsi}^{(2)}(\fett q) + \epsilon^3 F^{\rm (3a)}(\mD) \fett{\mPsi}^{\rm (3a)}(\fett q) \nonumber \\
&\qquad + \epsilon^3 F^{\rm (3b)}(\mD) \fett{\mPsi}^{\rm(3b)}(\fett q) +  \epsilon^3 F^{\rm (3c)}(\mD) \, \nab \times \fett {\mathscr{A}}^{\rm(3c)}(\fett q)   \label{eqs:LPTeps}
\end{align}
up to $O(\epsilon^4)$, where
\begin{align} 
 &\nab \cdot \fett \mPsi^{(1)} = \mu_1^{(1)} \,, \quad\, \nab \cdot \fett \mPsi^{(2)} = \mu_2^{(1,1)} \,,  \quad \,\, \nab \cdot \fett \mPsi^{\rm (3a)} = \mu_3^{(1,1,1)}\,, \nonumber \\ 
&\nab \cdot \fett \mPsi^{\rm (3b)} = \mu_2^{(1,2)} \,, \qquad\quad  \fett {\mathscr{A}}^{\rm(3c)} =  \nab \psi_{l}^{(1)} \times \nab \psi_{l}^{(2)} \,.  \label{eqs:Psi-vs-psi}
\end{align}
Within the context of standard LPT, the various spatial relationships~\eqref{eqs:Psi-vs-psi} arise as a consequence of exploiting the linearity of Poisson's equation, which implies that the full solution of the displacement at fixed order in $\epsilon$ can be constructed by means of a linear superposition \citep{1993MNRAS.264..375B,1994MNRAS.267..811B}.

\subsection{Resummation of $\Lambda$ terms in growth functions} \label{sec:resum}

Now we show how to determine the temporal coefficients in~\eqref{eqs:LPTeps}
within the context of our recursion relations~\eqref{eq:recs}. 
For this we set $G(\mD) \in \{ E, F^{\rm(3a)} , F^{\rm(3b)}, \ldots  \}$ with corresponding spatial displacement
$\fett \mPsi^{(G)}(\fett q) \in \{ \fett \mPsi^{(2)}, \fett \mPsi^{\rm(3a)}, \fett \mPsi^{\rm(3b)},  \ldots \}$,
as instructed through the expressions~\eqref{eqs:LPTeps}--\eqref{eqs:Psi-vs-psi},
and define  a functional derivative such that
\be \label{eq:funci}
 \boxed{  \left. \sdfrac{\delta\,  \fett \psi^{(n)}(\fett q)}{\delta \fett \mPsi^{(G)}(\fett q)} \right|_{\varphi^{\rm ini}\, =\,0} \equiv g_n \,,
 \qquad \text{with}~~\,\,   G(\mD) = \sum_{n=1}^\infty g_n \, D^n \,,  }
\ee 
where the $\fett{\psi}^{(n)}$'s are given by~\eqref{eq:recs}, and $g_n$ are coefficients that we want to determine. In words, the functional derivative~\eqref{eq:funci} extracts all terms with a certain spatial dependence from our recursion relations, while the additional requirement $\varphi^{\rm ini}=0$ ensures 
that only temporal coefficients with no remaining spatial dependence are selected.

To be specific, let us apply~\eqref{eq:funci} to extract $E(\mD)$ from our recursive relations. For this it
suffices to consider only the recursive relation for $\mu_1^{(n)}$ given in~\eqref{eq:recLongitudinal} and, using the identification~\eqref{eqs:Psi-vs-psi}, extract all terms that have spatial dependence $\mu_2^{(1,1)}$. Using the functional derivative
\be
   \left. \sdfrac{\delta\, \mu_1^{(n)}(\fett q)}{\delta\, \mu_2^{(1,1)}(\fett q)} \right|_{\varphi^{\rm ini} =0} = e_n \, ,
 \qquad \text{with}~~\,\,   E(\mD) = \sum_{n=1}^\infty e_n \, D^n \,,
\ee
and applying it to equation~\eqref{eq:recLongitudinal}, we find  $e_1 =0$ for $n=1$, and for $n \geq 2$, the 
simple recursive solutions
\be \label{eq:en}
 e_n = - \sdfrac 3 7 \delta_{2n} + \sum_{i+j=n}  \sdfrac{i(i-1)\,e_i \lambda^{(j)}}{ (n+3/2)(n-1)}  \,,
\ee
where the $\lambda^{(j)}$ coefficients are given through~\eqref{eq:lambda}. The first contributions for $E$ read
\be \label{eq:Egrowth}
  E(\mD) = -\sdfrac{3}{7} D^2 -\sdfrac{3 \Lambda}{1001}  D^5   -\sdfrac{960\Lambda^2}{3\,556\,553}  D^8 + O(D^{11})  \,. 
\ee
Of course, by construction, our functional derivative extracts the fastest-growing mode solution of the associated differential equation, which in the present case reads \citep{2021MNRAS.503..406R}
\be
  (1- \lambda) D^2 \pDsq E + \sdfrac{3D}{2} \pD E - \sdfrac{3}{2} E = - \sdfrac{3D^2}{2} \,.
\ee
We note that for the second-order growing mode, there actually exists an exact analytical solution expressed through combinations of hypergeometric functions \citep[see equation~22 of][]{1995PThPh..94.1151M}, and we have explicitly verified that our solutions coincide exactly to the 40th order in the Taylor expansion.
For higher-order growth functions beyond~$E$, however, we are not aware of exact analytical results in the literature.

Concerning higher-order growth functions, we find for  $\mathfrak{a}=$ a,b,c 
\be
  F^{\rm(3\mathfrak{a})}(\mD) = \sum_{n=1}^\infty f_n^{\rm(3\mathfrak{a})} \,D^n
\ee
the following simple recursive relations $(n \geq 3)$
\begin{align}
  & f_n^{\rm(3a)} = - \sdfrac 1 3 \delta_{3n} + \sum_{i+j=n} \sdfrac{i(i-1) \, f_i^{\rm(3a)} \lambda^{(j)}}{(n+3/2)(n-1)}  \,, \\ 
 & f_n^{\rm(3b)} = + \sdfrac 1 3 \delta_{3n} - \sdfrac{3e_{n-1}}{(n+3/2)(n-1)} + \sum_{i+j=n} \sdfrac{i(i-1)\,f_i^{\rm(3b)} \lambda^{(j)}}{(n+3/2)(n-1)}   \,, \\
  & f_n^{\rm(3c)} =  \left( 1- 2/n \right) e_{n-1} \,, 
\end{align}
where the coefficients $\lambda^{(j)}$ and $e_i$ are respectively given by equations~\eqref{eq:lambdas} and~\eqref{eq:en}. The first few contributions are
\begin{subequations} \label{eqs:Fabc}
\begin{align}
 &F^{\rm(3a)}(\mD) =  -\sdfrac {1} {3} D^3  - \sdfrac{4\Lambda }{825} D^6 - \sdfrac{109 \Lambda^2 }{215\,985} D^9 +O(D^{12}) \,,  \label{eq:Fagrowth} \\
&F^{\rm(3b)} (\mD) = \sdfrac{10}{21} D^3 + \sdfrac{538 \Lambda}{75\,075} D^6  +
 \sdfrac{3581 \Lambda^2}{4\,849\,845} D^9  +  O(D^{12}) \,, \label{eq:Fbgrowth} \\
&F^{\rm(3c)}(\mD) = - \sdfrac 1 7 D^3  - \sdfrac{2\Lambda}{1001} D^6 - \sdfrac{320\Lambda^2}{1\,524\,237} D^9 +  O(D^{12}) \label{eq:Fcgrowth}
\end{align}
\end{subequations}
and, as for the second-order growth, these results relate  to ordinary differential equations
which are respectively \citep{1995A&A...296..575B}
\begin{align}
  & (1- \lambda) D^2 \pDsq  F^{\rm(3a)} + \sdfrac{3D}{2} \pD  F^{\rm(3a)} - \sdfrac{3}{2}  F^{\rm(3a)} = - 3D^3  \,, \\
  & (1- \lambda) D^2 \pDsq  F^{\rm(3b)} + \sdfrac{3D}{2} \pD  F^{\rm(3b)} - \sdfrac{3}{2}  F^{\rm(3b)} =  3D^3 - 3 E D  \,, \\
  & \pD F^{\rm(3c)} = D \pD E - E \,. 
\end{align} 
These derivations carry on straightforwardly for higher-order growth functions.
We discuss our results below (see figures~\ref{fig:growth}--\ref{fig:spectra}), but before that we briefly address growth normalisation as well as truncation effects.

\subsection{Structure growth: Normalisation and truncation effects}\label{sec:norm-trunc}

So far we have not yet normalised the linear growth $D$ (cf.\ equation~\ref{eq:Dplus}).
Assuming current cosmological parameters, we have $D(a_0) =0.7801$ at $a_0=1$.  
To convert our results such that $D$ is unity  at the present time, we simultaneously rescale $D \to D/ D(a_0)$ and $\Lambda \to D^3(a_0) \Lambda$
in all our results for $E$, $F^{\rm(3a)}$, $F^{\rm(3b)}$, etc.
From here on we assume that such a rescaling has been already applied. 
See Appendix~\ref{app:disp-velocity} for a summary of our results, taking into account the standard normalisation for $D$.

\begin{table}
 \caption{Numerical values for the temporal coefficients $E$ and $F^{\rm(3a)}$ -- $F^{\rm(3c)}$ of the displacement~\eqref{eqs:LPTeps} evaluated today at $a=a_0$, as a function of truncation order $N$ in their infinite power series
  (equations~\ref{eq:Egrowth} and~\ref{eq:Fagrowth}--\ref{eq:Fcgrowth}).
 Shown results (see also Appendix~\ref{app:disp-velocity}) are normalised such that $D=1$ today ($z=0$).}
 \label{tab:convergencegrowth}
 \begin{tabular}{c|cccc}
  \hline
  truncation $N$ &  $E$ &  $F^{\rm(3a)}$ & $F^{\rm(3b)}$ & $F^{\rm(3c)}$  \\
   \hline
    3 & $-0.428571$ & $-0.333333$ & $0.476190$ & $-0.142857$\\
    6 & $-0.431861$ & $-0.338655$ & $0.484056$ & $-0.145050$\\
    9 & $-0.432186$ & $-0.339263$ & $0.484945$ & $-0.145303$\\
   12 & $-0.432237$ & $-0.339365$ & $0.485094$ & $-0.145345$\\
   15 & $-0.432247$ & $-0.339385$ & $0.485124$ & $-0.145354$\\
   18 & $-0.432249$ & $-0.339390$ & $0.485131$ & $-0.145356$\\
   21 & $-0.432250$ & $-0.339391$ & $0.485132$ & $-0.145357$\\
   24 & $-0.432250$ & $-0.339391$ & $0.485133$ & $-0.145357$\\ 
  \hline
 \end{tabular}
\end{table}

Next we briefly discuss the impact of truncation errors, i.e., errors
that occur in practice when only determining a finite number of terms
in the infinite power series for the growth functions, derived through~\eqref{eq:funci}.  
For this we show in table~\ref{tab:convergencegrowth} an overview of our predictions for the growth functions as a function of truncation order~$N$ in their power series
$E(\mD) = \sum_{n=1}^N e_n \, D^n$ and $ F^{\rm(3\mathfrak{a})}(\mD) = \sum_{n=1}^N f_n^{\rm(3\mathfrak{a})} \,D^n$ for $\mathfrak{a}\!=\!$ a,b,c,
evaluated at today's time~$D=1$ (this is the time when the speed of convergence is expected to be the slowest).
From the table it is seen that our results converge fast, and begin to be accurate to 4 [5] significant digits at truncation order~$N =12$~[$N=18$].
Henceforth we call our results ``exact'' if they are converged to at least~5~significant digits.

\subsection{Discussions on the displacement growth and velocity} \label{sec:discussgrowth}

\begin{figure}
 \centering
   \includegraphics[width=0.95\columnwidth]{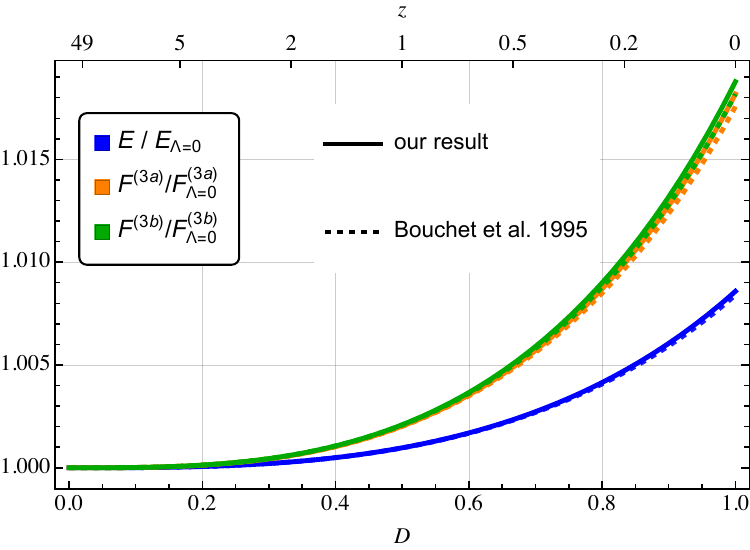}
   \caption{Temporal evolution of the growth functions $E$, $F^{\rm(3a)}$ and $F^{\rm(3b)}$ of the displacement~\eqref{eqs:LPTeps}, taking the standard normalisation of $D$ into account (i.e., also the $x$-axis is normalised). We show ratios of the growth functions over their predictions when all terms with explicit appearances of $\Lambda$ are set to zero (cf.\ equations~\ref{eq:Egrowth} and~\ref{eq:Fagrowth}--\ref{eq:Fcgrowth}).
    We include the first 20 Taylor coefficients in our results (solid lines), for which our results are converged to at least five significant digits (cf.\ table~\ref{tab:convergencegrowth}). Shown are also the asymptotic considerations of  \protect\cite{1995A&A...296..575B} (dashed lines), which agree with ours to better than $0.06\%$ precision. We note that the green dashed and orange solid lines are virtually overlapping.}   \label{fig:growth}
\end{figure}

In figure\,\ref{fig:growth} we show the temporal evolution of the displacement growth functions (solid lines)
based on equations~\eqref{eq:Egrowth},~\eqref{eq:Fagrowth} and~\eqref{eq:Fbgrowth}, and compare them against known results in the literature (dashed lines). Specifically, \cite{1995A&A...296..575B} provide leading-order asymptotic solutions for the growth functions (denoted with index ``B''),  which are
\begin{subequations} 
\begin{align}
 &D_{\mB}^{\phantom{a}} = D \,, \qquad E_\mB \asymp -\frac 3 7 \Omega^{1/143} D^2\,,  \label{eq:BouchetE}\\ 
 & F_\mB^{\rm(3a)}\asymp -\frac 1 3 \Omega^{-4/275} D^3 \,, \qquad F_\mB^{\rm(3b)}\asymp \frac{10}{21} \Omega^{-269/17875} D^3\,,
\end{align}
\end{subequations} 
where $\Omega = (1+ \Lambda a^3)^{-1}$.
From figure~\ref{fig:growth} it is seen that the effects of 
accurately incorporating $\Lambda$ for the considered growth functions
 are at most $1.9\%$ at late times, which seems to be in good agreement with \cite{2016JCAP...05..058T} (for $E$) and \cite{2021JCAP...04..033S} (for $F^{\rm(3a)}$ and $F^{\rm(3b)}$) who numerically integrated the  underlying differential equations. 
Furthermore, our solutions for the  growth functions agree to high precision with those of \cite{1995A&A...296..575B}, specifically to better than~$0.06\%$.

\begin{figure}
 \centering
   \includegraphics[width=0.95\columnwidth]{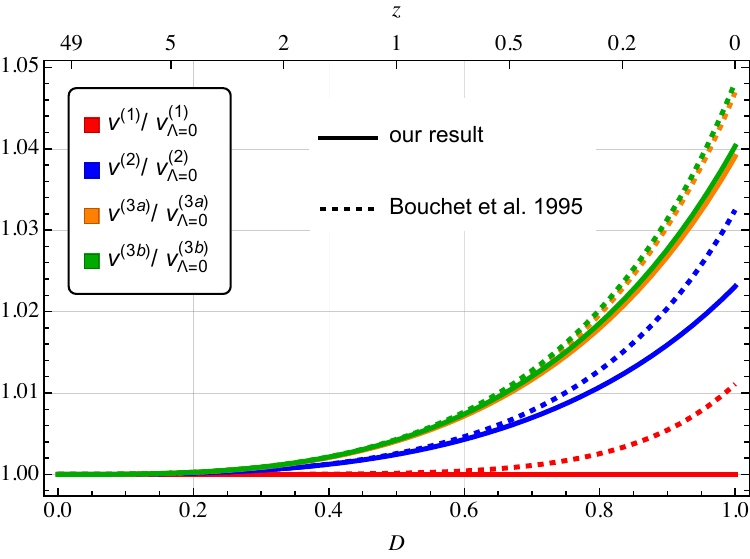}
   \caption{Same as previous figure but for the velocity coefficients~\eqref{eqs:velocitycoeffs}, taking the standard normalisation for $D$ into account (see also Appendix~\ref{app:disp-velocity}).}
  \label{fig:vel}
\end{figure}

Next we discuss the impact of $\Lambda$ on predictions of the velocity. For this, observe that the Lagrangian representation of the peculiar velocity
can be written as $\fett u = \partial_t \fett x = \dot D \fett v$, where we have defined $\fett v = \pD \fett \mPsi$.  Taking equation~\eqref{eqs:LPTeps} as the input for $\fett \mPsi$, we can write
\be \label{eq:v}
  \fett v 
   = \epsilon v^{(1)}  \fett{\mPsi}^{(1)}\!+ \epsilon^2 v^{(2)} \fett{\mPsi}^{(2)}\! + \epsilon^3 v^{\rm(3a)} \fett{\mPsi}^{\rm (3a)} \! 
  + \epsilon^3 v^{\rm(3b)} \fett{\mPsi}^{\rm(3b)} +  \ldots  \,,
\ee
where the various velocity coefficients $v^{(n)}$ follow directly from our solutions for the growths~\eqref{eq:Egrowth},~\eqref{eq:Fagrowth} and~\eqref{eq:Fbgrowth}; they are
\begin{subequations} \label{eqs:velocitycoeffs}
\begin{align}
   &v^{(1)} = 1 \,, \\
  &v^{(2)}  = -\sdfrac{6}{7} D -\sdfrac{15 \Lambda}{1001}  D^4   -\sdfrac{7680\Lambda^2}{3\,556\,553}  D^7 + O(D^{10}) \,,\\
  &v^{\rm(3a)} =  - D^2  - \sdfrac{8\Lambda}{275} D^5 - \sdfrac{327 \Lambda^2 }{71\,995} D^8  + O(D^{11}) \,, \\
  &v^{\rm(3b)} =   \sdfrac{10}{7} D^2 + \sdfrac{1076 \Lambda}{25\,025} D^5  +
 \sdfrac{10743 \Lambda^2}{1\,616\,615} D^8 + O(D^{11}) \,.
\end{align}
\end{subequations}
For a complete list of results that also take a normalised $D$ as input, see Appendix~\ref{app:disp-velocity}.
These results should be compared against the asymptotic considerations of \cite{1995A&A...296..575B} who find
\begin{subequations}
\begin{align}
 &v^{(1)}_\mB \asymp  \frac{D \Omega^{6/11}}{a \partial_a D} \,, \qquad  v^{(2)}_\mB  \asymp  \frac{2E_\mB \Omega^{153/286}}{a \partial_a D} \,, \\ 
  &v^{\rm(3a)}_\mB \asymp \frac{3F_\mB^{\rm(3a)} \Omega^{146/275}}{a \partial_a D}  \,, \qquad v^{\rm(3b)}_\mB \asymp \frac{3F_\mB^{\rm(3b)} \Omega^{9481/17875}}{a \partial_a D} \,.
\end{align}
\end{subequations}
In figure~\ref{fig:vel} we plot ratios of our solutions for the velocity, versus the predictions for setting $\Lambda=0$ when appearing explicitly in the expressions.
At late times, the ratios of velocities are of the order of a few per cent, which suggests that the $\Lambda$ corrections in the velocity are non-negligible. We also observe that the results of~\cite{1995A&A...296..575B}
show a slight trend of overestimating the effects of $\Lambda$CDM in the velocity (dashed lines in figure~\ref{fig:vel}). We were able to understand this mild deviation analytically: for example,
when expanding their first-order prediction of the velocity in powers of $D$, we find $v^{(1)}_\mB = 1 + (15 \Lambda^2/2057) D^6 + O(D^9)$ which deviates  from unity at late times, indicating that the asymptotic considerations of~\cite{1995A&A...296..575B} could be refined.
Furthermore, \cite{1995A&A...296..575B} provided also analytical fits for the velocity (their equations~50 and~51), however we observed a fairly similar trend in slightly over-predicting the amplitudes for the velocity growth, hence we do not show them here.

 Finally, it is well-known that peculiar motion can lead to a distortion in the observed position of matter particles \citep[e.g.][]{1987MNRAS.227....1K,2009MNRAS.393..297P}. Here  we briefly discuss the impact on these so-called redshift-space distortions when the explicit $\Lambda$ dependence in the matter growth is ignored.
For simplicity we limit ourselves to a Newtonian analysis as well as to the commonly used plane-parallel approximation, for which the relation between real-space position $\fett x$ relates to the position $\fett s$ in redshift space as follows \citep[see e.g.][]{2000ApJ...535....1M,2008PhRvD..77f3530M},
\be
  \fett s = \fett x + H^{-1}  u_z  \,\hat {\fett z} \,,
\ee
where $u_z =  \hat{\fett z} \cdot \fett u$ is the line-of-sight projection of the peculiar velocity $\fett u = \partial_t \fett x$. 
Now we consider the identity $\fett u = H f D \, \fett v$ where $f = \dd \ln  D / \dd \ln a$ is the logarithmic growth rate, and the rescaled velocity $\fett v$ is given in equation~\eqref{eq:v}.  From these relations it becomes clear that ignoring the explicit $\Lambda$ dependence in the higher-order structure growths in redshift space amounts to ignoring the explicit  $\Lambda$ dependence in $\fett v$  (while the time-dependent coefficient between $\fett u$ and $\fett v$ is kept unaltered). Hence, similar arguments as given above apply; in particular, figure~\ref{fig:vel} represents also the expected error on the position in redshift-space if the explicit $\Lambda$ dependence in the growth is discarded.

\subsection{Matter power spectrum and bispectrum} \label{sec:polyspectra}

So far we have provided our analytical results exclusively for the Lagrangian displacement field, which is the only dynamical variable in Lagrangian space. Here we show how the results in Lagrangian coordinates can be translated to predictions for the fluid variables in Eulerian coordinates.
In the following we outline the main steps and results, while calculational details are provided in Appendix~\ref{app:poly}.

Employing the  Dirac-delta $\delta_\mDirac^\mthree$,
the Eulerian density can be expressed using the displacement field   \citep[see e.g.][]{2008PhRvD..77f3530M},
\be \label{eq:-delta-dirac-delta}
 \delta(\fett x) + 1 = \int  \delta_\mDirac^\mthree(\fett{x} - \fett q - \fett \mPsi(\fett q)) \,\dd^3 q  \,, 
\ee
where from here on we occasionally drop some temporal dependence when there is no source of confusion.
Denoting by a tilde the Fourier conjugated variable to the Eulerian variable $\fett x$,
the  density contrast in Fourier space can be written as \citep{Taylor:1996ne}
\be \label{eq:Fourierdelta}
  \tilde \delta (\fett k) = \! \int\!\!\dd^3 \!q \,{\rm e}^{{\rm i} \ws \fett k \cdot\ws \fett q} \left[ {\rm e}^{\ii\ws \fett k \ws\!\cdot\!\ws \fett \mPsi(\fett q)} -1 \right]
  = \int\!\!\dd^3 \!q \,{\rm e}^{\ii \ws \fett k \!\ws\cdot\ws \fett q} \sum_{n=1}^\infty \sdfrac{(\ii\ws \fett k \!\cdot\! \fett \mPsi)^n}{n!} .
\ee
Now, replacing $\fett \mPsi =  \epsilon D  \fett{\mPsi}^{(1)} + \epsilon^2 E \fett{\mPsi}^{(2)} + \ldots$ 
in the last expression and collecting all terms with the same powers in $\epsilon$, we can write 
\be \label{eq:deltaexp}
 \tilde \delta(\fett k) =  \epsilon \tilde \delta_1(\fett k) + \epsilon^2 \tilde \delta_2(\fett k) + \epsilon^3 \tilde \delta_3(\fett k) + \ldots \,,
\ee
with
\begin{align}
 &\tilde \delta_n(\fett k,t) = \int \frac{\dd^3 k_{1} \cdots \dd^3 k_n}{(2\uppi)^{3n-1}} \delta_\mDirac^\mthree(\fett k - \fett k_{\!12\cdots n})\, F_n^{\rm (s)}(\fett k_1,\ldots, \fett k_n)\, \nonumber \\
 &\qquad \qquad \qquad \times \tilde \delta_1(\fett k_1,t) \cdots  \tilde \delta_1(\fett k_n,t) \,, 
  \label{eq:deltatildeFn}
\end{align}
where  $\fett{k}_{12\cdots n} = \fett k_1 + \fett k_2 + \ldots + \fett k_n$,
 and 
$\delta_1$ is the linear density contrast at time $t$. 
The first two kernels are $F_1^{\rm (s)} = 1$ and 
\begin{align}
  &F_2^{\rm (s)}
=  \sdfrac {1} {2}  + \sdfrac{3\cE}{14}  + \sdfrac{\fett k_1 \!\cdot \fett k_2}{k_1 k_2} \left[\sdfrac{k_1}{k_2} + \sdfrac{k_2}{k_1} \right] + \left[ \sdfrac{1}{2} - \sdfrac{3 \cE}{14}\right] \sdfrac{(\fett k_1 \! \cdot \fett k_2)^2}{k_1^2 k_2^2}  \label{eq:F2}
\end{align}
\citep{1995A&A...296..575B},
where $\cE = -7E/(3D^2)$.
In Appendix~\ref{app:poly} we also provide an expression for~$F_3^{\rm (s)}$ which is somewhat bulky.
Note that the $F_n^{\rm(s)}$ kernels are  generally time-dependent for a $\Lambda$CDM Universe,
but reduce to purely spatial relations in the limiting case of an EdS universe
 for which $\cE =1$; see e.g.\ \cite{Bernardeau2002}.

Next we determine perturbative versions for the matter powerspectrum $P$ and bispectrum $B$, defined respectively with
\begin{subequations}
\begin{align}
  &\langle \tilde \delta(\fett k_1)\, \tilde \delta(\fett k_2)  \rangle = (2\uppi)^3 \delta_\mDirac^\mthree(\fett{k}_{\!12})\, P(k_1)  \,,\\
  &\langle \tilde \delta(\fett k_1)\, \tilde \delta(\fett k_2)  \, \tilde \delta(\fett k_3)  \rangle = (2\uppi)^3 \delta_\mDirac^\mthree(\fett k_{\!123})\, B(k_1, k_2, k_3) \,,
\end{align}
where $k_n = |\fett k_n|$. Plugging in the perturbative expansion~\eqref{eq:deltaexp} in the left-hand-sides of 
these definitions and assuming Gaussian initial conditions, one gets \citep{Bernardeau2002}
\begin{align}
  &P =  P_{11}\, \epsilon^2 +  P_{\rm loop} \, \epsilon^4  + O(\epsilon^6) \,, \quad P_{\text{loop}}  = P_{22}  +  P_{13}\,,  \\
  &B = B_{211}\,\epsilon^4 + O(\epsilon^6) \,,
\end{align}
 \end{subequations}
where $P_{ij}(k_1) (2\uppi)^3 \delta_\mDirac^\mthree(\fett{k}_{\!12}) := \langle \tilde \delta_i(\fett k_1) \tilde \delta_j(\fett k_2)\rangle$ and analogously for the bispectrum. Here, $P_{11}$ is the linear matter power spectrum that can be obtained from conventional Einstein--Boltzmann solvers.
After a few straightforward calculational steps, derived in Appendix~\ref{app:poly}, one obtains
\begin{align}
  &P_{22}(k,t) = \frac{k^3}{98 (2\uppi)^2}  \int_0^\infty \dd r \,P_{11} (k r,t)   \int_{-1}^{+1} {\rm d} x \,P_{11} (K,t) \nonumber \\
    &\qquad \quad \times \frac{(3 \cE r + 7x - [7+ 3 \cE ] r x^2)^2}{(1+r^2 - 2 r x)^2} , \label{eq:P22} \\
 & P_{13}(k,t) = \frac{k^3 P_{11}(k,t)}{252 (2\uppi)^2} \int_0^\infty  \dd r P_{11} (k r,t) \bigg[ 3(9 \cE - 5 \cF)r^{-2} - 168 \nonumber \\
 &\quad - 45 \cE + 55 \cF   + ( 45 \cE + 55 \cF ) r^2   - 3 ( 9 \cE + 5 \cF ) r^4 \nonumber \\
&\quad + \frac{3}{2r^3} (r^2 -1)^3 ( [9 \cE + 5 \cF] r^2 +9 \cE - 5 \cF) \ln \left| \frac{1+r}{1-r} \right| \bigg] \label{eq:P13} ,
\end{align}
where $K =k\sqrt{1+r^2 - 2rx}$ 
 and $\cF = 21F^{\rm(3b)}/(10D^3)$.
Ignoring all $\Lambda$ dependences implies setting $\cE = 1 = \cF$ (the EdS model), 
for which the above results 
coincide exactly with those given by~\cite{1991PhRvL..66..264S}.
We remark that, starting from a similar calculation, \cite{2014PhRvD..89h4017L} and 
\cite{2014PhLB..736..403L} determined also the one-loop power spectrum (using a resummation technique; see also \citealt{2008PhRvD..77f3530M}), 
with the temporal coefficients being calculated numerically.
However,  we were unable to reproduce their numerical results.
In particular, we do obtain substantially different predictions for the $\Lambda$CDM growth functions, which appears to be the source of the discrepancy  (see e.g.\ figure~3 of \citealt{2014PhRvD..89h4017L}).

For the tree-level contribution to the matter bispectrum, occasionally denoted with $B_{211}$, one finds
\be
  B = 2 P_{11}(k_1,t)\, P_{11}(k_2,t) \,F_2^{\rm (s)}(\fett k_1 ,\fett k_2) + \text{two~perms.} \,,
\ee
which is a well-documented expression in the literature (e.g.\ \citealt{1998ApJ...496..586S}), but, to our knowledge,  is usually evaluated
within an approximation that amounts to ignoring the explicit $\Lambda$ terms in $E$ (cf.\ equation~\ref{eq:Egrowth}); 
see e.g.~\cite{1995A&A...298..643H} for an exception employing the asymptotic solutions~\eqref{eq:BouchetE}.

\begin{figure}
 \centering
   \includegraphics[width=0.95\columnwidth]{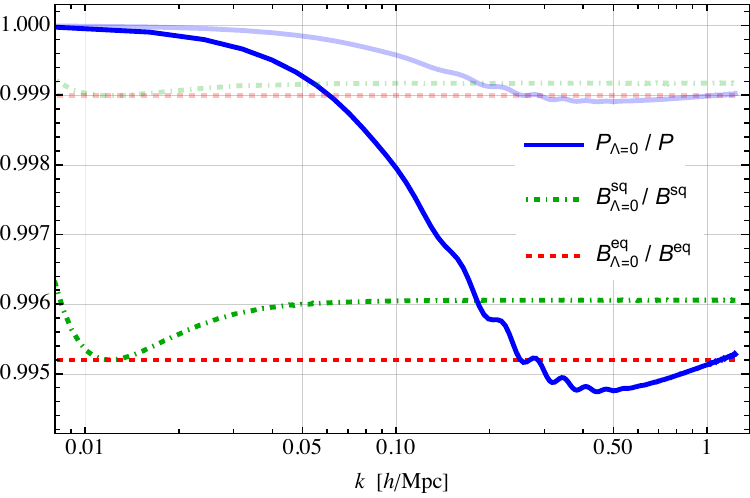}
   \caption{Ratios of various statistics showing the effect of ignoring the $\Lambda$~terms in the growth functions, evaluated today at redshift $z=0$ [faint lines at $z=1.0$]. Specifically, results shown in blue (solid) 
are for the total matter power spectrum up to one-loop accuracy, whereas the green dot-dashed and red-dashed lines refer to the tree-level matter bispectrum for  squeezed and equilateral triangle configurations, respectively.}   \label{fig:spectra}
\end{figure}

In figure~\ref{fig:spectra} we show 
ratios of the total power spectrum up to one-loop accuracy when $\Lambda$ is switched off/on in the second- and third-order structure growths (respectively equations~\ref{eq:Egrowth} and~\ref{eqs:Fabc}). The blue solid [faint] line is evaluated at $z=0$ [$z=1.0$]. 
This figure indicates that  ignoring the explicit $\Lambda$ contributions in the growth lead to a maximal power suppression of about $0.53\%$ at $k \simeq 0.45 h/$Mpc for $z=0$ [$0.11\%$ at $k \simeq 0.45 h/$Mpc for $z=1.0$], which is, admittedly, a mild effect.
However, the shown scale-dependent power suppression has a fairly similar signature known from massive neutrino cosmologies
\citep[see e.g.][]{2012MNRAS.420.2551B,2020JCAP...09..018P,2020ApJ...904..159Y,2020arXiv201112504Z}. 
Using the fit $\Delta P / P_{\rm max} \simeq - 9.8 \,\Omega_{\nu0}/ (\Omega_{\rm m0} - \Omega_{\nu0})$ of \cite{2008JCAP...08..020B} for the non-linear power suppression  due to massive neutrinos with current density $\Omega_{\nu0} = \sum {\rm m}_\nu/(93.14\,{\rm eV} h^2)$, we estimate that the effect of employing inaccurate $\Lambda$CDM growth functions within a Markov chain Monte Carlo likelihood analysis
could lead to systematic errors of the order of $\Delta \sum {\rm m}_\nu \approx 0.0073\,$eV on predicting the sum of massive neutrinos, assuming a normal mass hierarchy of neutrinos.

Actually, that we observe such a non-linear power suppression in the present case is not surprising, and has been already observed by the (numerical) approaches of e.g.~\cite{2006MNRAS.366..547M,2008JCAP...10..036P,2014JCAP...11..039B,2021JCAP...01..020G}: As in massive neutrino cosmologies, the dominant effect on the matter dynamics comes from an updated Hubble drag term in the equations of motion~\eqref{eq:EOMbasic}, which itself is the consequence of an altered Friedmann equation. We have seen that accurately incorporating~$\Lambda$ in the Hubble drag leads to updated perturbative results for the displacement at essentially all orders. Thus, as expected, the updated Hubble drag leads to a nonlinear backreaction to the motion of matter elements.

In figure~\ref{fig:spectra} we also show the power suppression due to~$\Lambda$ at the example of the  leading-order contribution to the matter bispectrum in two distinguished triangle configurations, specifically  for squeezed triangles with $k_1 = k_2 = k$ and $k_3 = \Delta k =0.012 h/$Mpc (``sq''; green dot-dashed lines), as well as for equilateral triangles with $k_1 = k_2 = k_3 = k$ (``eq''; red dashed lines).
In the squeezed case we observe a very mild scale dependence near $\Delta k$, otherwise the power suppressions in the shown bispectra are essentially constant and of order $0.5\%$ at $z=0$ [order $0.1\%$ at $z=1$].
These results for the tree-level matter bispectrum seem to be in good agreement with the
analytical considerations of~\cite{1995A&A...298..643H} who employed the asymptotic solutions of \cite{1995A&A...296..575B}, as well as with the numerical computations of \cite{2021PhRvD.104l3551B}.
Again, these are fairly mild effects for accurately incorporating $\Lambda$. None the less, we encourage to employ accurate growth functions in order to further reduce uncertainties in the theoretical modelling.

Finally, as elucidated in section~\ref{sec:discussgrowth}, we expect that ignoring the $\Lambda$-dependence in the matter velocity leads to an altered position of visible matter in redshift-space. It would be interesting to analyse in  detail related redshift-space observables, such as the  power- and bispectrum in redshift-space, while employing our refined matter growth and velocity. We leave such avenues for future work.

\section{Summary and concluding remarks} \label{sec:concl}

{\bf Summary.}
We have derived all-order recursive relations~\eqref{eq:recs} for the Lagrangian displacement field in a $\Lambda$CDM Universe. Similar recursive relations have been already reported in the literature \citep{2012JCAP...12..004R,2014JFM...749..404Z,2015MNRAS.452.1421R,2015PhRvD..92b3534M,2021JCAP...04..033S}, but the novelty of the present approach is at least twofold. Firstly, we do not discard any couplings between $\Lambda$ and CDM as it is commonly done in the literature. Secondly, our recursive relations are of purely spatial character, since the temporal evolution has been fully integrated out. This is made possible by a reformulation of the temporal dependence in the cosmic fluid equations from  $t$-time to $D$-time (equations~\ref{eq:master}).
The temporal part of the reformulated equations can then be explicitly solved by a power-series {\it Ansatz} in~$D$, from which we obtain the said recursive solutions for the displacement.

After performing suitable resummation techniques, described in section~\ref{sec:resum}, we show how the results obtained in the strict $D$ expansion can be converted to results within the framework of standard (Lagrangian) perturbation theory, where the initial gravitational potential can be viewed as the perturbative bookkeeping parameter. The output of the resummation  constitutes  analytical growth functions for the displacement (equations~\ref{eq:Egrowth} and~\ref{eqs:Fabc}), which are expressed in terms of fast converging  expressions in powers of $D$ (table~\ref{tab:convergencegrowth}).

We demonstrate our resummation method by specifically determining accurate displacement growth functions at second- and third-order in perturbation theory -- commonly denoted with $E$ and $F$, respectively. The temporal evolution of these growth functions and their first time derivatives (i.e the temporal velocity coefficients) are respectively shown in figures~\ref{fig:growth} and~\ref{fig:vel}, and compared against the leading-order asymptotic considerations of~\cite{1995A&A...296..575B}. While our results for $E$ and $F$ agree very well with the predictions of~\cite{1995A&A...296..575B}, we report a mild disagreement of order $1$\% at late times for all considered velocity coefficients. In general, the effect of accurately incorporating $\Lambda$ in the velocity coefficients is up to $4$\% at the considered orders, and thus should be taken into account especially for accurate forward modelling at late times.

The so obtained resummed results for the  displacement and velocity can be directly applied to generate highly accurate 3LPT initial conditions for $N$-body simulations; for a concise summary of results see the Appendix~\ref{app:disp-velocity}, and we have also updated the publicly available initial condition generator \texttt{monofonIC} \citep{2020ascl.soft08024H,2021MNRAS.500..663M} accordingly.

Furthermore, our results for the structure growth in Lagrangian coordinates can be easily translated to predictions for the Eulerian density field, essentially by exploiting the exact relationship~\eqref{eq:-delta-dirac-delta}. 
The corresponding perturbation kernels for the density in Fourier space can be used to determine the impact of~$\Lambda$ on  the one-loop matter power spectrum and on the tree-level matter bispectrum (see e.g.\ \citealt{1995A&A...298..643H,2006MNRAS.366..547M,2008JCAP...10..036P,2014JCAP...11..039B,2021JCAP...01..020G,2021PhRvD.104l3551B} for related avenues).
The results are summarised in figure~\ref{fig:spectra} where we show ratios for the total polyspectra when $\Lambda$ is switched off and on.
Again, the $\Lambda$ effects are fairly mild but, at the same time, are easily incorporated, e.g.\ by modifying existing codes \citep[e.g.][]{2012PhRvD..86j3528T,2017JCAP...02..030F,2018JCAP...04..030S,2021JCAP...03..100C} for determining loop corrections to polyspectra in real- or redshift space. \\

\noindent {\bf Concluding remarks.}
Whilst in the present work we focused on a standard $\Lambda$CDM cosmology, our techniques can be straightforwardly applied, or extended, to more generic cosmological models. For example, our results are also valid for (simplified) models of two or more cold fluids \citep{2021MNRAS.503..426H}, provided one updates the boundary conditions as instructed by  \cite{2021MNRAS.503..406R}.

Our methodology can  also accommodate for non-trivial modifications at the background level. This in particular allows for the inclusion of a spatial background curvature term, as well as certain (parametrisations  of) dark energy models \citep{2003MNRAS.346..573L,2015APh....63...23H,2020EPJC...80.1210V}. At the fluctuation level, the corresponding linear growth function ${\cal D}$  will look vastly different for non-standard cosmologies \citep[see e.g.][]{2005GReGr..37.2063D} as compared to the expression for $\Lambda$CDM.
However, as long as the respective modifications in the Friedmann equations can be represented in terms of (infinite) sums of polynomials in the scale factor with lowest exponent larger than $-3$, it follows that the altered growth~${\cal D}$ will be representable in terms of a Taylor series in~$a$ or, vice versa, in terms of powers of its inverse function~$a(\text{\small ${\cal D}$})$. Hence, recursive relations in powers of~${\cal D}$ can be obtained, from which would follow the resummed growth of structure for certain non-standard cosmologies.

The case of massive-neutrino cosmology deserves special attention, as in principle this requires a description that  incorporates at least the leading-order contributions from general (and special) relativity. However, \cite{2020JCAP...09..018P,2022arXiv220113186H} have shown that the non-linear gravitational feedback of massive neutrinos on matter can be effectively incorporated by a suitable choice of coordinates \citep[this is the so-called Newtonian motion gauge framework, see e.g.][]{2016JCAP...09..031F,2017JCAP...06..043F}. Within the weak-field limit of general 
relativity, the resulting relativistic equations of motion for matter take then precisely the Newtonian form, which are  equations~\eqref{eq:EOMbasic}, the starting point of our approach. Hence, our results for the higher-order growth functions directly apply to massive neutrino cosmologies,
provided that the sum of neutrino mass is below $0.3$\,eV; 
see \cite{2022arXiv220113186H} for details. 
We remark that the gravitational feedback of massive neutrinos on matter can also be incorporated  through a refined linear growth function, see \cite{2022arXiv220200670E} for an explicit implementation employing updated LPT recursive relations.

Given our refined growth functions, let us comment on the time of shell-crossing and the radius of convergence of the perturbative series in Lagrangian space.
The former is the instant when the fluid velocity becomes multi-valued and the density singular, thereby marking the physical breakdown of the single-stream description. 
 \cite{2021MNRAS.501L..71R} have determined the time of shell-crossing for cosmological initial conditions, however using recursive relations that amount to ignore all explicit $\Lambda$ dependences in the structure growths. While at first sight this seems to be a crude approximation, it should be noted that for $\Lambda$CDM initial conditions, the estimated time of first shell-crossing is, roughly, at redshifts $z \approx 10-25$ for a particle Nyquist frequency $k_{\rm Ny} = \uppi N^{1/3}/L$ of around $k_{\rm Ny} = ( 25 - 5 ) h/$Mpc, where $N$ is the number of particles and $L$ the box size of the numerical simulation (cf. figure~A.3 of \citealt{2021MNRAS.501L..71R}).
At such high redshifts, the impact of $\Lambda$ on the structure growth is however minuscule; for example, the $\Lambda$ correction for the second- and third-order growth functions is of order $\sim 10^{-5}$ at $z=10$.
Similar statements can be made for the associated time when LPT convergence is lost, which, for realistic initial conditions and equivalent resolutions as above, occurs at redshifts $z\approx 8 - 21$ (see table~1 of~\citealt{2021MNRAS.501L..71R}).

Of course, the above just tells us (again) that ignoring the explicit~$\Lambda$ dependence in the growth functions  becomes relevant only at fairly late redshifts.
At the same time, physically appropriate solutions that exclude shell-crossings can be obtained by applying suitable filtering techniques to the initial data, such as performed within the effective theory for the cosmic large scale-structure \citep{2012JCAP...07..051B,2012JHEP...09..082C,2021JCAP...04..033S}. The resulting coarse-grained, effective fluid can then be evolved very accurately using our refined growths, while the gravitational feedback stemming from small-scale physics (baryons, post-shell-crossing physics, etc.; see e.g. \citealt{2015JCAP...05..019L,2017MNRAS.470.4858T,2021MNRAS.505L..90R}) may be incorporated in the analysis by suitable counter terms within the effective description.

\section*{Acknowledgements}

We thank Thomas Montandon for useful comments on the manuscript.
O.H.\ acknowledges funding from the European Research Council (ERC) under the European Union’s Horizon 2020 research and innovation programme, Grant Agreement No.~679145 (COSMO-SIMS).

\section*{Note added}

After completing  this work,  \cite{2022arXiv220510026F} presented a similar analysis for the $\Lambda$CDM growth functions, focusing mostly on the computation of the matter power spectrum. While the approaches are different in nature, the magnitude of the considered effects seem to agree reasonably well.

\section*{Data Availability}

There are no new data associated with this article.
The code to generate 3LPT initial conditions using the refined growth functions is freely  available at \url{https://bitbucket.org/ohahn/monofonic}.


\bibliographystyle{mnras}
\bibliography{biblio} 


%
\appendix

\section{Normalised displacement field and velocity in ready-to-use form}\label{app:disp-velocity}

For the sake of compactness, our results in the main text for the displacement and velocity growth 
are shown for unnormalised structure growth functions. Here we provide our main results employing the normalised growth function $\hat D(a) := D(a)/D_0$, where $D_0 := D(a_0)$ with $a_0$ denoting the present time, and $D(a)$ is the {\it unnormalised growth as defined} in~\eqref{eq:Dplus}.\footnote{If instead  an unnormalised growth $\tilde D = c D$ is used, where $D$ is given in~\eqref{eq:Dplus} and $c$ is an arbitrary parameter, then for each $D_0$ appearing in equations~\eqref{eq:Egrowthlong}--\eqref{eq:v3c}, the following rescaling should be performed $D_0 \to D_0/c$.} We remind the reader that $\Lambda = \Omega_{\Lambda0}/\Omega_{\rm m0}$.

The normalised growth functions for the displacement series
\begin{align}
  &{\fett{\mPsi}} = \epsilon \hat D  \fett{\mPsi}^{(1)}(\fett q)+ \epsilon^2 \hat E \fett{\mPsi}^{(2)}(\fett q) + \epsilon^3 \hat F^{\rm (3a)} \fett{\mPsi}^{\rm (3a)}(\fett q) \nonumber \\
&\qquad + \epsilon^3 \hat F^{\rm (3b)} \fett{\mPsi}^{\rm(3b)}(\fett q) +  \epsilon^3 \hat F^{\rm (3c)} \, \nab \times \fett {\mathscr{A}}^{\rm(3c)}(\fett q)  
\end{align}
is
\begin{align}
 &  \hat E = -\sdfrac{3}{7} \hat D^2 -\sdfrac{3 \Lambda}{1001}  D_0^3 \hat D^5   -\sdfrac{960\Lambda^2}{3556553} D_0^6  \hat D^8  -\sdfrac{2040  \Lambda^3}{52929877} D_0^9 {\hat D}^{11}     \nonumber \\
 & \quad -\sdfrac{1038060000 \Lambda^4}{151269407005717} D_0^{12} {\hat D}^{14}
   -\sdfrac{85671376080 \Lambda^5}{61566648651326819}  D_0^{15} {\hat D}^{17} \nonumber \\
 & \quad- \sdfrac{1864100579256192 \Lambda^6}{6062846087823581445961} D_0^{18} \hat D^{20}  + O(\hat D^{23}) \,, \label{eq:Egrowthlong}   
\end{align}
\begin{align}
 & \hat F^{\rm(3a)} =  -\sdfrac{{\hat D}^3}{3}
-\sdfrac{4\Lambda }{825} D_0^3 {\hat D}^6 
-\sdfrac{109  \Lambda^2}{215985} D_0^6 {\hat D}^9
-\sdfrac{15408  \Lambda^3}{200362085} D_0^9 {\hat D}^{12} \nonumber \\
& \quad-\sdfrac{46316624 \Lambda^4}{3259690760865} D_0^{12} {\hat D}^{15} 
-\sdfrac{4677016576  \Lambda^5}{1584861647932563} D_0^{15} {\hat D}^{18} \nonumber \\
&\quad-\sdfrac{330321967776  \Lambda^6}{498174844666802303} D_0^{18} {\hat D}^{21}
 + O(\hat D^{24})  \label{eq:Falong}
  \,, 
\end{align}
\begin{align}
 & \hat F^{\rm(3b)}  = \sdfrac{10 {\hat D}^3}{21}
  +\sdfrac{538  \Lambda}{75075} D_0^3 {\hat D}^6
  +\sdfrac{3581  \Lambda ^2}{4849845} D_0^6 {\hat D}^9 \nonumber \\
 & \quad+\sdfrac{16644976  \Lambda^3}{148468304985} D_0^9 {\hat D}^{12}
+\sdfrac{24575717136 \Lambda^4}{1188545340759205} D_0^{12} {\hat D}^{15}  \nonumber \\
& \quad+\sdfrac{11397154716512  \Lambda^5}{2656837684107257343} D_0^{15} {\hat D}^{18} \nonumber \\
& \quad+\sdfrac{1348899757328270624  \Lambda^6}{1400517446287247314016991} D_0^{18} {\hat D}^{21}
+ O(\hat D^{24})
\end{align}
\begin{align}
 & \hat F^{\rm(3c)}  =  -\sdfrac{{\hat D}^3}{7}
 -\sdfrac{2  \Lambda }{1001} D_0^3 {\hat D}^6
-\sdfrac{320  \Lambda^2}{1524237} D_0^6 {\hat D}^9 \nonumber \\
& \quad-\sdfrac{1700  \Lambda^3}{52929877} D_0^9 {\hat D}^{12}
-\sdfrac{69204000  \Lambda ^4}{11636108231209} D_0^{12} {\hat D}^{15} \nonumber \\
&\quad-\sdfrac{228457002880  \Lambda^5}{184699945953980457} D_0^{15} {\hat D}^{18} \nonumber \\
&\quad-\sdfrac{621366859752064  \Lambda^6}{2233680137619214216933} D_0^{18} {\hat D}^{21} + O(\hat D^{24}) \,.
\end{align}
For the velocity 
\begin{align}
  &\fett v = \partial_D \fett \mPsi
   = \epsilon v^{(1)}  \fett{\mPsi}^{(1)}\!+ \epsilon^2 v^{(2)} \fett{\mPsi}^{(2)}\! + \epsilon^3 v^{\rm(3a)} \fett{\mPsi}^{\rm (3a)} \! 
  + \epsilon^3 v^{\rm(3b)} \fett{\mPsi}^{\rm(3b)}  \nonumber \\
 &\quad +  \epsilon^3 v^{\rm (3c)}(\mD) \, \nab \times \fett {\mathscr{A}}^{\rm(3c)}(\fett q) \,,
\end{align}
the coefficients are
\begin{align}
   &v^{(1)} = 1 \,, 
\end{align}
\begin{align}
  &v^{(2)}  =  -\sdfrac{6 {\hat D}}{7} 
 -\sdfrac{15  \Lambda }{1001} D_0^3 {\hat D}^4
 -\sdfrac{7680 \Lambda^2}{3556553} D_0^6 {\hat D}^7
 -\sdfrac{2040  \Lambda^3}{4811807} D_0^9 {\hat D}^{10} \nonumber \\
 &\quad  -\sdfrac{2076120000  \Lambda ^4}{21609915286531} D_0^{12} {\hat D}^{13}
 -\sdfrac{85671376080  \Lambda ^5}{3621567567725107} D_0^{15} {\hat D}^{16}  \nonumber \\
 &\quad  -\sdfrac{37282011585123840  \Lambda ^6}{6062846087823581445961} D_0^{18} {\hat D}^{19}
  + O(\hat D^{22})  \,,
\end{align}
\begin{align}
  &v^{\rm(3a)} =   -{\hat D}^2 
 -\sdfrac{8\Lambda }{275} D_0^3 {\hat D}^5 
 -\sdfrac{327  \Lambda^2}{71995} D_0^6 {\hat D}^8
 -\sdfrac{184896  \Lambda ^3}{200362085} D_0^9 {\hat D}^{11}  \nonumber \\
 & \quad -\sdfrac{46316624  \Lambda ^4}{217312717391} D_0^{12} {\hat D}^{14}
 -\sdfrac{28062099456 \Lambda^5}{528287215977521} D_0^{15} {\hat D}^{17}  \nonumber \\
 & \quad  -\sdfrac{990965903328  \Lambda^6}{71167834952400329} D_0^{18} {\hat D}^{20}
  + O(\hat D^{23})  \,, 
\end{align}
\begin{align}
  &v^{\rm(3b)} =   \sdfrac{10 {\hat D}^2}{7}
 +\sdfrac{1076  \Lambda}{25025} D_0^3 {\hat D}^5
 +\sdfrac{10743  \Lambda^2}{1616615} D_0^6 {\hat D}^8  \nonumber \\
 & \quad  +\sdfrac{66579904 \Lambda^3}{49489434995} D_0^9 {\hat D}^{11} 
 +\sdfrac{73727151408  \Lambda^4}{237709068151841} D_0^{12} {\hat D}^{14}  \nonumber \\
 & \quad +\sdfrac{68382928299072  \Lambda^5}{885612561369085781} D_0^{15} {\hat D}^{17}  \nonumber \\
 & \quad  +\sdfrac{1348899757328270624  \Lambda^6}{66691306966059395905571} D_0^{18} {\hat D}^{20}
 + O(\hat D^{23})   \,, 
\end{align}
\begin{align}
  &v^{\rm(3c)} =  -\sdfrac{3 {\hat D}^2}{7}
 -\sdfrac{12 \Lambda}{1001} D_0^3 {\hat D}^5 
 -\sdfrac{960  \Lambda^2}{508079} D_0^6 {\hat D}^8
 -\sdfrac{20400  \Lambda^3}{52929877} D_0^9 {\hat D}^{11}  \nonumber \\
 & \quad  -\sdfrac{1038060000  \Lambda^4}{11636108231209} D_0^{12} {\hat D}^{14}
 -\sdfrac{1370742017280 \Lambda^5}{61566648651326819} D_0^{15} {\hat D}^{17}  \nonumber \\
 & \quad  -\sdfrac{1864100579256192  \Lambda^6}{319097162517030602419} D_0^{18} {\hat D}^{20}
  + O({\hat D}^{23}) \,. \label{eq:v3c}
\end{align}
Finally, we remark that the above rescaled velocity $\fett v$ relates to the peculiar velocity $\fett u = \partial_t \fett x$ as follows,
\be
  \fett u = H f \hat D \, \fett v \,,
\ee where $H$ is the Hubble parameter, and $f = \dd \ln \hat D / \dd \ln a$ is the logarithmic derivative of the linear growth factor. Finally, for completeness, we also provide $f$ in terms of the normalised growth,
\begin{align}
  f &= 1 - \sdfrac{6 \Lambda}{11} D_0^3  \hat D^3 +  \sdfrac{240\Lambda^2}{2057} D_0^6  \hat D^6 
      -  \sdfrac{1200\Lambda^3}{47311} D_0^9  \hat D^9  \nonumber \\
  &\qquad   + \sdfrac{15382080\Lambda^4}{2822243083} D_0^{12}  \hat D^{12} 
     -  \sdfrac{36723360\Lambda^5}{31044673913} D_0^{15}  \hat D^{15}  \nonumber \\
  &\qquad   +   \sdfrac{126547707648\Lambda^6}{497677167499303} D_0^{18}  \hat D^{18}  + O({\hat D}^{21}) \,.
\end{align}

\section{Calculational details to  derivation of one-loop power spectrum}\label{app:poly}

Here we provide calculational details to the derivations that lead to the one-loop matter powerspectrum (equations~\ref{eq:P22}--\ref{eq:P13}).

Starting from the $\Lambda$CDM results for the displacement and using the identity~\eqref{eq:Fourierdelta} for the Eulerian density, one obtains the perturbation kernels $F_n^{\rm (s)}$ as defined for equation~\eqref{eq:deltatildeFn}. The first few perturbation kernels, already in symmetrised form, read
\begin{align}
 &F_1^{\rm (s)}(\fett k_1) = 1 \,, \label{eq:F1} 
\end{align}
\begin{align}
  &F_2^{\rm (s)}(\fett k_1,\fett k_2) = \sdfrac {1} {2}  + \sdfrac{3\cE}{14}  + \sdfrac{\fett k_1 \!\cdot \fett k_2}{k_1 k_2} \left[\! \sdfrac{k_1}{k_2} \! +\! \sdfrac{k_2}{k_1} \! \right] \!+ \! \left[ \sdfrac{1}{2} - \sdfrac{3 \cE}{14}\right] \sdfrac{(\fett k_1 \! \cdot \fett k_2)^2}{k_1^2 k_2^2} , \label{eq:F2rep} 
\end{align}
\begin{align}
  &F_3^{\rm (s)}(\fett k_1,\fett k_2, \fett k_3) =   \bigg\{ \sdfrac{5\cF}{42} 
  \left[ 1- \sdfrac{(\fett k_2 \cdot \fett k_3)^2}{k_2^2 k_3^2}\right] \left[ 1- \sdfrac{(\fett k_1 \cdot \fett k_{23})^2}{k_1^2 k_{23}^2} \right] \nonumber  \\
  &\qquad +   \cE \sdfrac{\fett k_{123} \cdot \fett k_1}{k_1^2} \sdfrac{\fett k_{123} \cdot  \fett k_{23}}{14k_{23}^2} \left[ \sdfrac{(\fett k_2 \cdot \fett k_3)^2}{k_2^2 k_3^2} -1 \right]+ \text{two perms.}\bigg\} \nonumber\\
 &\qquad   + \frac 1 6 \sdfrac{\fett k_{123} \cdot \fett k_1}{k_1^2} \, \sdfrac{ \fett k_{123} \cdot \fett k_2}{k_2^2}  \,  \sdfrac{\fett k_{123} \cdot \fett k_3}{k_3^2}  +  \fett k_{123} \cdot \fett S^{\rm (3a)} \,,
\end{align}
where $\cE = -7E/(3D^2)$ and $\cF = 21F^{\rm(3b)}/(10D^3)$, while $\fett S^{\rm (3a)}$ is a scalar contribution $\propto \mu_3^{(1,1,1)}$ that however does not contribute to the one-loop correction of the matter power spectrum, hence we do not include it here (but see e.g.\ equation~2.13 of \citealt{2012JCAP...06..018R}). 

Using these kernels for equation~\eqref{eq:deltaexp} and considering standard perturbative expansions of
the two-point  Fourier-space correlator $\langle  \tilde \delta(\fett k_1)\, \tilde \delta(\fett k_2)  \rangle$, it is elementary to arrive at  
\begin{align}\label{eq:power-oneloop}
 &P(k,t) = P_{11}(k,t) + P_{\rm loop}(k,t)\, , \\ 
 & P_{\rm loop}(k,t) = P_{22}(k,t) + P_{13}(k,t)
\end{align}
for Gaussian initial conditions,
with the perturbative correlators
\begin{align}\label{eq:Pij}
  \left\langle \tilde\delta_i(\fett{k}_1,t)  \,\tilde\delta_j(\fett{k}_2,t)  \right\rangle := (2\uppi)^3 \delta_{\rm D}^\mthree \left( \fett{k}_{12} \right) P_{ij}(k_1,t) \,,
\end{align}
where $P_{11}$ is the linear matter power spectrum, while
\begin{align}
 & P_{22}(k) = 2 \! \int \! \frac{\dd^3p}{(2\uppi)^3} [F_2^{\rm (s)}(\fett{k}-\fett{p},\fett{p})]^2 P_{11}(|\fett{k}-\fett{p}|)\,P_{11}(p) \, , \label{eq:P22rep} \\
 &  P_{13}(k) = 6 P_{11}(k) \! \int \! \frac{\dd^3p}{(2\uppi)^3} F_3^{\rm (s)}(\fett{k}, \fett{p},-\fett{p})  P_{11}(p) \, , \label{eq:P13rep}
\end{align}
are the first non-vanishing loop integrals for Gaussian initial conditions \citep{Bernardeau2002}.
Upon moving from Cartesian to spherical coordinates, some of the angular integrations for equations~\eqref{eq:P22rep}--\eqref{eq:P13rep} can be performed explicitly (cf.\ \citealt{1991PhRvL..66..264S}), which straightforwardly leads to equations~\eqref{eq:P22}--\eqref{eq:P13}.

For the residual numerical integration of the one-loop power spectra, we utilise the deterministic integration routine \texttt{CUHRE} from the \texttt{CUBA} library \citep{2005CoPhC.168...78H}.

\bsp	
\label{lastpage}
\end{document}